\documentclass[prb,twocolumn,showpacs,amsmath,amssymb]{revtex4-1}
\usepackage{graphicx} 
\usepackage{epsfig}
\usepackage{dcolumn} 
\usepackage{color}
\usepackage{graphicx}
\usepackage{amsmath,amssymb}
\begin{document}

\title{Law without law or
``just'' limit theorems? \\ \small{~\\ Some reflections about a proposal of Wheeler's}}
\author{Sergio Caprara$^{1,2,*}$ and Angelo Vulpiani$^{1,3,*}$}
\affiliation{$^1$Dipartimento di Fisica - Universit\`a di Roma Sapienza, Piazzale Aldo Moro 5 - 00185 Roma, Italy \\
$^2$Institute for Complex Systems (ISC-CNR), UOS Sapienza, Piazzale A. Moro 5, 00185 Roma, Italy\\
$^3$Centro Interdisciplinare  Linceo ``B. Segre'', Accademia dei Lincei,
Via della Lungara, 10 - 00165 Roma, Italy \\
$^*$Correspondence and requests for materials should be addressed \\ to S.C. (sergio.caprara@roma1.infn.it) 
or to A.V. (angelo.vulpiani@roma1.infn.it)}

\begin{abstract}
About $35$ years ago  Wheeler introduced the motto ``law without
law'' to highlight the possibility that  (at least  a part of) Physics
may be understood only following {\em regularity principles} and few
{\em relevant facts}, rather than relying on a  treatment in
terms of fundamental theories. 
Such a proposal can be seen as  part of a more general attempt (including 
the maximum entropy approach) summarized by the slogan ``it from bit'', 
which privileges the information as the basic ingredient. 
Apparently it seems that it is possible to obtain, without
the use of physical laws, some important results in an easy way,
for instance, the probability distribution of the canonical ensemble. 
In  this paper we will present a general discussion on those ideas of Wheeler's that 
originated the motto ``law without law''. In particular we will show how the
claimed simplicity is only apparent and it is rather easy to produce wrong results.
We will show that it is possible to obtain
some of the  results  treated by Wheeler in the realm of the statistical 
mechanics, using precise assumptions and nontrivial results of probability 
theory, mainly concerning ergodicity and limit theorems.
\end{abstract}

\maketitle

\section{Introduction}
\label{intro}
In an influential paper \cite{uno} reporting his Oersted Medal Response at the joint APS-AAPT Meeting 
(25 January 1983) John Archibald Wheeler posed the following bold problem: Is it possible 
to build part of Physics only following {\em regularity principles} (RP)?
In other  words, can we understand (at least a part of) Physics without referring to fundamental theories 
(e.g., classical or quantum mechanics)?

Wheeler proposed the ``law without law'' program. 
For instance, in Ref.\,[\onlinecite{uno}] he discussed three examples which,
in his opinion, support the approach in terms of RP:
\\
(i) the Boltzmann-Gibbs distribution of the canonical ensemble; 
\\
(ii) the universality of critical exponents; 
\\
(iii) the statistical features of heavy nuclei. 
\\
Regarding point (i),  Wheeler wonders: ``How can stupid molecules ever be conceived to obey a 
law so simple and so general? [...] What regulating principle
accomplished this miracle?''. And then he adds: ``No small answer can ever hope to live up to a question 
so great''.

According to Wheeler \cite{uno}, Physics may be divided into three Eras:
\\
$\bullet$  Era number one (Copernicus, Kepler, and Galileo) is characterized
by the discovery of the simplicity of motion;
\\
$\bullet$ Era number two (starting with Newton) is marked by the concept of physical laws;
\\
$\bullet$ Era number three (present time) is ruled by regulating principles, i.e., ``chaos behind law''. 

Here, the basic idea is that chaotic behavior and regulating principles can fruitfully cooperate and give rise to 
approximated laws. Let us note that Wheeler uses the term ``chaos'' in a  wide sense, i.e., without
the technical meaning of sensitivity to the initial conditions.

His message can be summarized in the motto
\begin{eqnarray}
&&\Big( higgledy- piggledy \Big) +\Big( regulating \,\,  principle\Big)=\nonumber\\
&&\Big( law \,\, of \,\,  physics \Big) \, .
\label{one}
\end{eqnarray}

Wheeler's proposal can be seen as a part of a more general program whose famous 
slogan  is ``it from bit'', suggesting that the ultimate building block upon which we should 
base our knowledge of the World is information \cite{due}.

Deutsch \cite{tre}, discussing Wheeler's idea,  puts forward 
the following dilemma: are the {\em regulating principles} analytic or synthetic propositions? 
He then tried to rewrite Wheeler's proposal in the less inspiring form
\begin{eqnarray}
&&\Big( stochastic \,\, laws\Big)  \to\nonumber\\ 
&&\Big( approximate \,\, ``deterministic \,\, laws"\Big)\, .
\label{two}
\end{eqnarray} 
Surely the proposal by Wheeler, even in the weaker Deutsch' form, is quite vague, on the other it 
is somehow part of a general approach to physics where the main ingredients are information and 
inference. Our main aim is a critical discussion of the limits of such an approach.

In particular, we will show that the examples discussed by Wheeler can appear as consequences
of general regulatory principles only a posteriori.
Actually  they follow from results of probability theory that are far from trivial, 
basically the main ingredients are limit theorems, namely:
\\
$\bullet$  The Boltzmann Gibbs law follows from the microcanonical distribution, which is based on 
(some form of) the ergodic hypothesis.
\\
$\bullet$  The universality of the statistical features of heavy nuclei is a sort of ergodicity: the single 
(large) nucleus is well described by its average properties.
\\
$\bullet$ The universal properties of critical phenomena can be seen as a generalization of the central 
limit theorems.

In Sec.\,II  we address the issue of the Boltzmann-Gibbs distribution, discussing an enlightening conjecture 
by Ulam, the notion of ``bridge law'', the ergodic hypothesis and Khinchin's approach, and the microcanonical 
distribution.
In Sec.\,III we cast our reflections about Wheeler's proposal into a wider perspective, discussing the case of
universality in critical phenomena. In Sec.\,IV we address the role and meaning of the maximum entropy principle, 
comment upon the actual role of probability in statistical physics.
Some final remarks are found in Sec.\,V.

\section{The Boltzmann-Gibbs distribution of the canonical ensemble} 

Let us  briefly discuss some general aspects
of the foundations of the statistical mechanics, in particular the role of ergodicity and of the many degrees 
of freedom in macroscopic objects \cite{quattro,cinque,sei}. 

We owe to Ludwig Boltzmann's ingenuity two great ideas in modern Physics \cite{cinque,sette}: 
\\
(A) the introduction of probability in the  physical realm;
\\
(B) the bridge between the microscopic dynamics, and the macroscopic description of the physical world, which 
is based on a phenomenological theory, named thermodynamics.

Before discussing in some detail Boltzmann's approach, we open a brief digression about an interesting conjecture 
by Ulam, which helps to clarify some aspects about statistical laws in physics,
in particular the unavoidable role of the dynamics, or at least some aspects
of the laws ruling the time evolution.

\subsection{An interlude about Stupid Molecules and Boltzmann's law}

In his paper [\onlinecite{uno}], Wheeler writes:
{\it How can stupid molecules ever be conceived to obey a law so simple and so general? [...] What regulating 
principle accomplished this miracle?} And then he adds ``{\it No small answer can ever hope to live up to a 
question so great}''.
He discusses a model of $N$ oscillators sharing quanta of energy, wondering how in the limit $N \gg 1$ 
{\it one see how stupid molecules, sharing energy higgledy-piggledy, nevertheless end on the average obeying
Boltzmann's law?} He then concludes that this in an example of ``law without law'' according to 
[\onlinecite{uno}].

Let us briefly discuss an interesting conjecture of Ulam's.
Consider a (large) system with $N$ particles and the following stochastic rule:
at time $t$ a pair  $(i,j)$ is selected at random, these two particles perform a 
``collision'' which preserves the total energy but allows for a redistribution of the
energy according to the rule:
\begin{eqnarray*}
E_i(t+1)&=&X(t) \Big[ E_i(t) +E_j(t) \Big],\nonumber\\
E_j(t+1)&=&\left[1-X(t)\right] \Big[ E_i(t) +E_j(t) \Big],\nonumber
\end{eqnarray*}
where $X(t)$ is a random variable independent of $X(t-1)$, and takes values in $[0,1]$.
Ulam conjectured that, if $X$ is uniformly distributed, starting from any initial distribution of the 
energy, asymptotically Boltzmann's probability density for the energy holds:
$P_E(E)=\alpha \mathrm e^{-\alpha E}$ where $\alpha=1/E_0$ is determined by the 
mean energy  $E_0$  per particle at $t=0$.

\begin{figure*}
\begin{center}
\includegraphics[width=10cm]{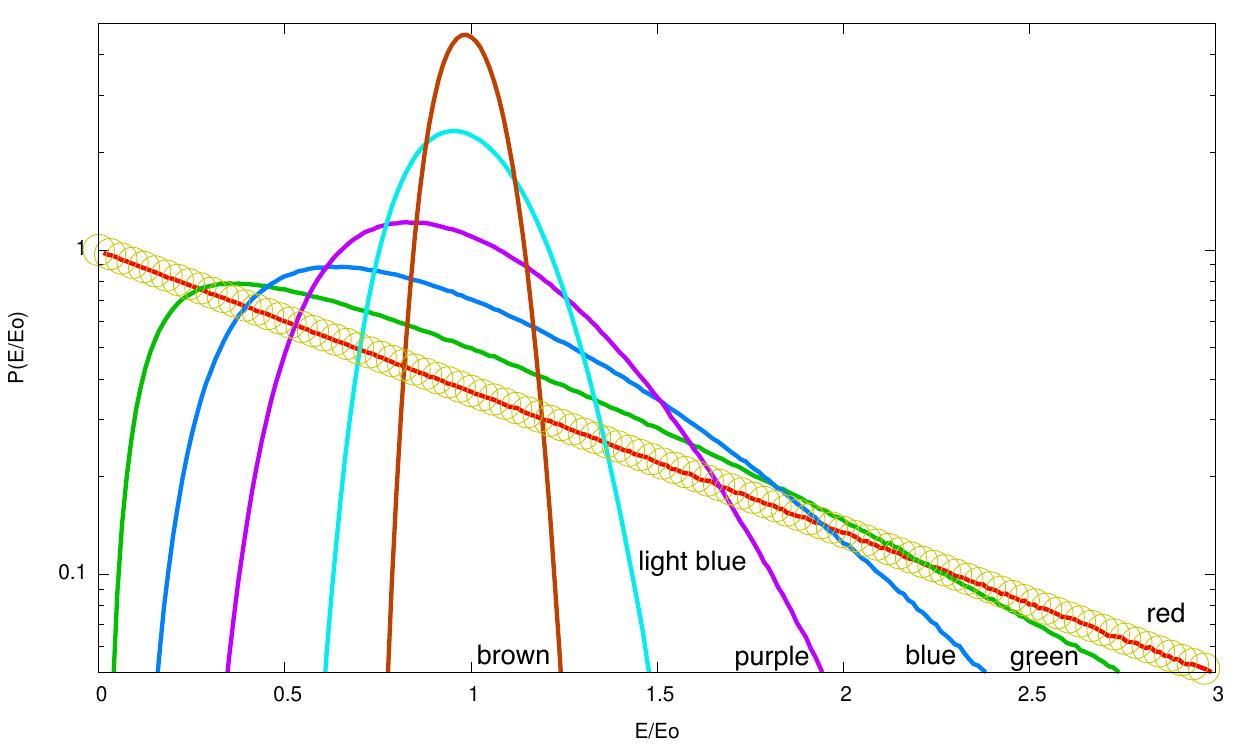}\\
\includegraphics[width=10cm]{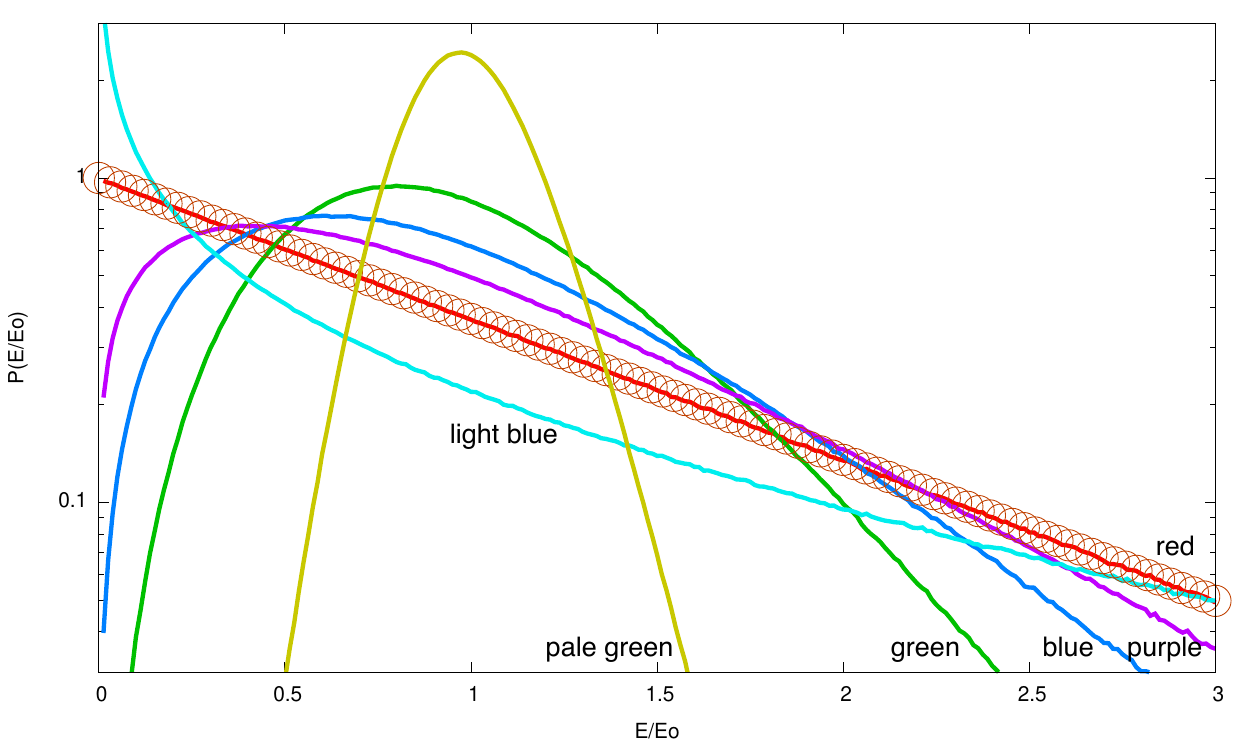}
\end{center}
\caption{
Top panel: Probability distribution of the energy $P_E(E/E_0)$, for various distributions densities $P_X(X)$
which are constant over a given interval $I\subset [0:1]$, and zero outside it:
red line, $I=[0:1]$; green line, $I=[0.1:0.9]$; blue line, $I=[0.2:0.8]$; purple line,
$I=[0.3:0.7]$; light blue line, $I=[0.4:0.6]$; brown line, $I=[0.45:0.55]$. Boltzmann's probability 
density $\mathrm e^{-E/E_0}$ is highlighted by the empty circles. 
Bottom panel:
Probability distribution of the energy $P_E(E/E_0)$, for various symmetric nonuniform distributions densities 
$P_X(X)=P_X(1-X)$, 
$X\subset[0:1]$ (the expression for $X\le\frac{1}{2}$ is given): pale green line, $P_X(X)=10(2X)^9$; 
green line, $P_X(X)=3(2X)^2$; blue line, $P_X(X)=4X$; purple line, $P_X(X)=\frac{3}{2}\sqrt{2X}$; light blue line, 
$P_X(X)=1/(2\sqrt{2X})$; the result for a uniform distribution is reported for comparison (red 
line). Again, Boltzmann's probability density $\mathrm e^{-E/E_0}$ is highlighted by the empty circles.}
\label{fig-ulam}
\end{figure*}
The above conjecture has been proved in a rigorous way by Blackwell and Mauldin \cite{otto}.
In addition it is possible to  generalize the result to the case of a generic
probability distribution $P_X(X)$: independently of the initial distribution of the energy, asymptotically 
one has a convergence to an invariant probability density for the energy $P_E(E)$ whose 
shape depends on $P_X(X)$. For instance, if $P_X(X)=\delta(X -\frac{1}{2})$, using the law of large numbers 
it is possible to show that $P_E(E)=\delta(E-E_0)$ where $E_0$ is the mean  energy per particle at $t=0$.
In Fig.\,\ref{fig-ulam} we report numerical calculations obtained for a
system with $N=50000$ particles, undergoing $500000$ collisions. The last (asymptotic) distribution is 
smoothened by averaging over the last $500$ realizations.

The results in Ref.\,[\onlinecite{otto}], as well as the above simple numerical computations, show in 
an unambiguous way the relevance of the ``correct rule''. 
Only for a specific $P_X(X)$ one obtains the proper physical results:
it is necessary to use a suitable stochastic law, while 
a generic ``higgledy-piggledy rule'' is not enough. The different results with varying $P_X(X)$ are somehow 
related to Bertrand's well-known paradox in probability theory \cite{bbb}: 
the probability $p$ that, considering an equilateral triangle inscribed in a circle, and picking at random a 
chord of the circle, the chosen chord is longer than a side of the triangle, depends on the way we interpret 
the prescription ``at random''. For instance, 
if the two end points of the chord are uniformly distributed along the circumference, this probability 
is $p=\frac{1}{3}$;
if the midpoint of the chord is uniformly distributed along a randomly chosen radius, $p=\frac{1}{2}$; 
if the midpoint of the 
chord is uniformly distributed over the entire circle, $p=\frac{1}{4}$. From these results one concludes 
that the term
``at random'' is ambiguous and therefore a precise prescription is needed to compute $p$.

We close this interlude noting that
it is possible to reformulate Ulam's approach in terms of velocity.
A particularly  interesting case is
given the granular gases where in the inelastic collision between two granular particles
the momentum is preserved and in addition each particle interacts with a thermal bath.
At variance with the elastic case the probability distribution of the velocity is not Gaussian
and shows long tails \cite{puglio} whose details depend on the parameters of the
collision rules, e.g. the restitution coefficient and the characteristic time of the
interaction particle-bath.

\subsection{Bridge law}
The issue (B) can be summarized by the well-known relation
\begin{equation}
\mathcal S =\kappa\log \mathcal W,\label{entropy}
\end{equation}
which is engraved on Boltzmann's tombstone in Vienna.
Actually the above equation, usually called {\it Boltzmann's law},
has been written by Planck \cite{sette}. 

Being  $H{({\mathbf Q}, \mathbf P)}$  the Hamiltonian of  a given system
whose particles are  in a volume  $V$, 
 $\mathcal S$ is the entropy (a thermodynamic quantity), 
\begin{equation}
\mathcal W(E,V,N)=\frac{1}{N!h^{3N}}\int_{E< H{({\mathbf Q}, \mathbf P)}<E+ \Delta E}\mathrm d^{3N} \mathbf Q \,
\mathrm d^{3N} \mathbf P\label{phase}
\end{equation}
is the ``volume of phase space'' (the space of all the coordinates $\mathbf Q$ and the momenta 
$\mathbf P$ specifying a 
microscopic configuration) accessible to a macroscopic configuration of $N$ particles with energy 
in the interval $[E, E + \Delta E]$, 
$\kappa$ is a dimensional constant (the Boltzmann constant) and 
$h$ is another dimensional constant (the Planck constant).

A simple, yet important, remark to the {\em bridge law} is in order here. Often, even in good textbooks 
(see, e.g., Ref.\,[\onlinecite{nove}]), one can read that the bridge law between thermodynamics and 
mechanics is given by the equation
\[
\langle \,\mathrm{kinetic ~energy~ per~ particle}\,\rangle=\frac{3}{2}\kappa T,
\]
relating the average kinetic energy of a particle with the absolute temperature $T$. This relation, which 
was already known 
to Daniel Bernoulli, holds in systems whose Hamiltonian contains the usual kinetic term 
$\sum_k \mathbf P_k^2/2m$.
On the other hand it cannot be valid for a system with a generic Hamiltonian. As an important example we 
can mention those systems which can have negative temperatures, e.g., point-vortex systems in 
two-dimensional inviscid fluids \cite{dieci}.

Eq. (\ref{entropy}), with $\mathcal W(E,V,N)$ given by Eq.  (\ref{phase}),
is the true {\em bridge law}, joining mechanics, i.e. $\mathcal W(E,V,N)$,
and thermodynamics, i.e. $\mathcal S (E,V,N)$.
Once $\mathcal W(E,V,N)$ is known, we have the entropy and therefore 
the temperature $T$ and the pressure $P$ can be obtained through
\[
\frac{1}{T}=\left(\frac{\partial \mathcal S}{\partial E}\right)_{V,N};~~~~~~~~
\frac{P}{T}=\left(\frac{\partial \mathcal S}{\partial V}\right)_{E,N}.
\]

\subsection{The ergodic hypothesis}

The issue (A) is rather subtle, and is still an open question. To shed light on this problem,
let us consider a macroscopic system with $N$ interacting particles, whose microscopic state
is described by the vector $\mathbf X\equiv(\mathbf Q_1,...,\mathbf Q_N;\mathbf P_1,...,\mathbf P_N)\in 
\mathbb R^{6N}$.
When an instrument measures a physical observable quantity, e.g., the pressure, it performs a time average of some
function of $\mathbf X$,
\begin{equation}
\frac{1}{\mathcal T}\int_0^{\mathcal T} \mathcal A(\mathbf X(t))\,\mathrm dt,\label{time}
\end{equation}
over the observation time $\mathcal T$. Of course, in order to determine the time evolution $\mathbf X(t)$, 
the knowledge of the initial condition $\mathbf X(t=0)$ is needed. However, even in this case, the program 
can be accomplished only for systems composed of a reasonably small number of particles and is definitely 
hopeless when $N$ grows as large as the Avogadro number 
(the number of H$_2$ molecules contained in just 2 grams of molecular hydrogen). 

Boltzmann's ingenious idea consists in replacing the time average in Eq. (\ref{time}) with a suitable 
average over phase space, assuming that
\begin{equation}
\lim_{\mathcal T\to\infty}\frac{1}{\mathcal T}\int_0^{\mathcal T} \mathcal A(\mathbf X(t))\,\mathrm dt=
\int \mathcal A(\mathbf X)\rho_{micro}(\mathbf X)\,\mathrm d\mathbf X,\label{ensemble}
\end{equation}
where  $\rho_{micro}(\mathbf X)$ is the  microcanonical probability density.
The limit $\mathcal T\to \infty$ is necessary from a mathematical point of view \cite{quattro,cinque}
and it is physically well justified. The reason is that the usual  observation time is much larger than the 
typical molecular times, e.g., the mean collision time in gases and liquids, or the oscillation periods 
in solids, which are $O(10^{-10} \,\mathrm s)$.
Once Eq. (\ref{ensemble}) is assumed to be valid in the limit of large $N$, in the presence of short-range
interactions, it is 
straightforward to derive, for a system exchanging energy with a much larger external environment, the 
Boltzmann-Gibbs distribution for the canonical ensemble
\[
\rho_{can}(\mathbf X)=\mathrm{Const.}\,\mathrm e^{-\beta H(\mathbf X)},
\]
with $\beta\equiv1/(\kappa T)$.

Now, after the Fermi-Pasta-Ulam simulations \cite{sei,dieci} and the  Kolmogorov-Arnol'd-Moser 
theorem \cite{undici}, 
it is quite clear that, strictly speaking, the EH cannot be valid.
On the other hand, in one of the most common and powerful method used in the numerical computation,
the molecular dynamics \cite{dodici}, the time average of observables are computed assuming 
the validity of (\ref{ensemble}).

Why, in spite of the failure of the ergodicity, the results
obtained with the Boltzmann-Gibbs distribution are valid?
In particular why is it in agreement with the results of molecular dynamics? 
Let us briefly discuss the physical reasons which allow for the practical
success of the standard computational techniques used in statistical mechanics.

Following Khinchin's approach \cite{tredici}, it can be
show that the EH is essentially true if (i) the system under study consists of a large number of particles, (ii) 
only ``suitable'' observable quantities are considered, and (iii) we allow for Eq. ({\ref{ensemble}}) to fail in 
a ``sufficiently small region'' (i.e., with a small probability). Khinchin was indeed able to show that for 
{\em sum functions} like
\begin{equation}
f(\mathbf X)=\sum_{k=1}^N f_k(\mathbf Q_k,\mathbf P_k)\label{obs}
\end{equation}
the inequality
\[
\mathrm{Probability}\,\left( |\delta f(\mathbf X)|\ge\frac{c_1}{N^{1/4}}\right)\le \frac{c_2}{N^{1/4}}
\]
holds, where $\delta f(\mathbf X)$ is the difference between the time average starting from $\mathbf X$ 
and the ensemble average, and $c_1,c_2$ are numerical constants. 
In the above equation the probability of the event $ |\delta f({\bf X})|\ge  c_1 N^{-1/4}$ is computed 
according to the microcanonical distribution.

The result, originally obtained by Khinchin  for systems of non interacting particles, and then  extended by 
Mazur and van der Linden \cite{mazur},
to systems of particles interacting through a short-range potential, 
can be summarized as follows:
although the EH does not hold in a rigorous mathematical sense, it is still
``physically'' valid provided
\\
a)  we  are interested only in suitable observables, with the shape (\ref{obs});
\\
b) we are  a bit ``tolerant''  
and  accept that ergodicity holds $O(N^{-1/4})$, i.e.,
the difference between the time average and the ensemble average is $O(N^{-1/4})$;
\\
c) ergodicity can fail in regions whose probability is as small
as $O(N^{-1/4})$, which vanishes for $N\to\infty$.
\\
Khinchin's result is a good example of what is usually called an 
{\em emergent property}: the dynamics has a marginal role, 
in the sense that in this context the crucial ingredient is the large number of particles 
$N$, being related to a technical aspect of probability theory, i.e., the 
{\em Law of large numbers} \cite{quattordici}.
Of course the details of the dynamics can have also a rather important role, e.g., 
this is a quite clear in Ulam's model discussed in Sec.\,2.1. Basically in all the problems of 
non equilibrium statistical physics it is not possible to avoid a treatment in terms of the underlying 
dynamics, see, e.g., the many delicate topics in the celebrated Fermi-Pasta-Ulam problems \cite{ggg}.

\subsection{On the microcanonical distribution}

Let us open a short digression about the microcanonical distribution $\rho_{micro}({\bf X})$ which is
constant if $H({\bf X}) \in [E, E+\Delta E]$ and zero otherwise.

Such an assumption sometimes is justified with the following argument:
since we cannot be able to have a control of a complex system with a large number of components,
we assume that probability distribution is uniform in the allowed region.
The above reasoning is not convincing: consider the variable ${\bf X}$ with a uniform
probability density $\rho_X({\bf X})$ in a certain region, then the probability density 
$\rho_Y({\bf Y})$ of another
variable ${\bf Y}={\bf F}({\bf X})$,  in general is not constant, therefore a uniform distribution,
at variance with some folklore, has no special status among the possible distributions;
in the following we will discuss again this point.

In our opinion the true physical good reason to adopt the microcanonical distribution is the following:
$\rho_{micro}({\bf X})$ is a stationary solution of the Liouville's equation.
We can say that the dynamics, somehow, enters in the selection of the probability. 

\section{Critical phenomena}
The second case discussed by Wheeler can be understood in a similar way: in critical phenomena, 
the universal behavior 
near the critical point is explained by means of the Renormalization Group, which has a probabilistic 
interpretation in terms
of generalized central limit theorems for non independent aleatory variables \cite{quindici}. 

It is well known that, for the validity of the central limit theorem, two necessary conditions must 
be met: the random variables must have a finite variance and be uncorrelated (or weakly correlated). 
Assume that the random
variables $x_1,...,x_n$ are independent and identically distributed with $p(x)$ having zero mean and 
$\sigma<+\infty$,
the variable $y_n=\frac{1}{\sigma\sqrt{n}}(x_1+...+x_n)$ is distributed according to a normalized 
Gaussian distribution. In the
case $p(x)\sim |x|^{-\alpha}$, with $1<\alpha<3$, $\sigma=+\infty$ and we have another limit theorem: the variable 
$y_n=\frac{1}{n^\beta}(x_1+...+x_n)$, with $\beta=1/(\alpha-1)$, for $n\gg 1$, is distributed according 
to the so-called
L\'evy stable distribution $\mathcal L_\alpha(x)$, and at large $|x|$ one has  
$\mathcal L_\alpha(x)\sim |x|^{-\alpha}$.
As an explicit example, where it is possible to write  $\mathcal L_2(x)$, we can cite the case of the
Cauchy-Lorentz distribution $P_\gamma(x)=\frac{\gamma}{\pi}(x^2+\gamma^2)^{-1}$, whose 
variance is infinite and in this case, for large $n$, the sample average $\frac{1}{n}(x_1+...+x_n)$ is 
not distributed according 
to a Gaussian, rather it follows the Cauchy-Lorentz distribution with exactly the same parameter $\gamma$ 
as the individual 
variables. Thus, averaging Cauchy-Lorentz-distributed variables does not lead to a narrower distribution.

The ubiquitous occurrence of the Gaussian distribution in statistical physics is ultimately due to the fact 
that correlations 
are usually short-ranged. However, approaching the critical point that marks a second order phase transition, 
with varying a control
parameter like the temperature, the microscopic degrees of freedom become correlated on larger and larger 
scales. As a result, 
very different physical systems manifest striking similarities in their nearly critical behavior.
For instance, several physical quantities tend to diverge (or vanish) as a power law when approaching the phase 
transition point, and different systems are characterized by the very same set of critical exponents (the numbers 
that characterize the power-law divergence, or vanishing, of various physical quantities). All 
the systems sharing the 
same set of critical exponents are said to belong to the same {\em class of universality} \cite{sedici}. 
In Wheeler's perspective, 
the question might be cast in the following manner: How can very different physical systems ``know'' 
that they will share 
the same set of critical exponents? It is now clear that the universality of critical phenomena 
can be understood in terms 
of deviations from the central limit theorem, and a new class of limit theorems in probability 
theory is indeed required 
to properly describe the properties of a physical system near a second order phase transition.

The idea underlying critical phenomena is that the variables that provide a microscopic description 
of a given system 
become more and more correlated when approaching a phase transition, so that the conditions that 
guarantee that the macroscopic variables, obtained by averaging over the microscopic degrees of freedom, 
follow a Gaussian distribution (as a consequence of the central limit theorem) are violated. The 
renormalization group approach provides the technical tool
to derive the asymptotic distributions.
\begin{figure*}
\includegraphics[width=12cm]{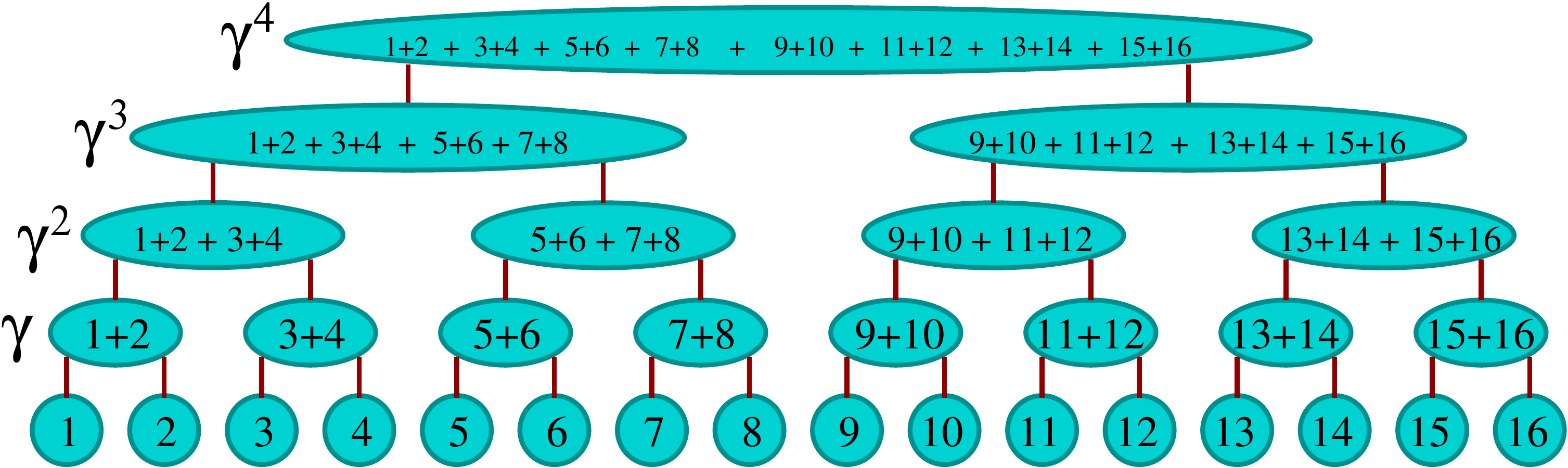}
\caption{Example of a hierarchical model with $N=4$.}
\label{fig-ger}
\end{figure*}
A prototypical model \cite{diciassette} that illustrates this increasingly strong correlation 
is described by a Hamiltonian 
defined on a set
of $2^N$ variables $x_i$, obeying the following hierarchical structure (see also Fig.\,\ref{fig-ger}):
\begin{eqnarray}
H_k(x_1,...,x_{2^k})&=&H_{k-1}(x_1,...,x_{2^{k-1}})\nonumber\\
&+&H_{k-1}(x_{2^{k-1}+1},...,x_{2^k})\nonumber\\
&-&\gamma^k\left(\sum_{i=1}^{2^k}
\frac{x_i}{2}\right)^2,
\label{hier}
\end{eqnarray}
with $k=1,...,N$, $H_0=0$ and the coupling $\gamma$ satisfying $1<\gamma<2$ (for $\gamma<1$ 
the hierarchical interaction 
scales to zero and the system is trivial, for $\gamma>2$ the system is thermodynamically unstable). Let us assume 
for simplicity that the distribution of the individual $x_i$ is Gaussian:
\[
p_0(x)=\frac{1}{\sqrt{2\pi}}\,\mathrm e^{-x^2/2},
\] 
with unit variance. In this case, starting from the Gibbs distribution of $2^N$ variables,
\[
\mathrm e^{-\beta H_N(x_1,...,x_{2^N})}\prod_{i=1}^{2^N} p_0(x_i),
\]
we can analytically proceed to regroup variables into blocks (lets call $\mathcal R$ 
the action of this regrouping procedure
on the distribution) and obtain a recursion relation for the distribution of the {\em block variables}
\begin{equation}
x^{(k)}=\sum_{i=1}^{2^k}\frac{x_i}{2^k}.
\label{block1}
\end{equation}
After $k$ iterations one has
\[
p_k(x)\equiv\left(\mathcal R^k p_0\right)(x)=\frac{1}{\sqrt{2\pi\sigma_k^2}}\,\mathrm e^{-x^2/(2\sigma_k^2)},
\]
with variance
\[
\sigma_k^2=\frac{1}{1-2\beta\sum_{\ell=1}^k\left(\frac{\gamma}{2}\right)^\ell}.
\]
For $k\to\infty$ the limiting distribution is Gaussian,
\[
p_\infty(x)=\frac{1}{\sqrt{2\pi\sigma_\infty^2}}\,\mathrm e^{-x^2/(2\sigma_\infty^2)},
\]
with variance
\[
\sigma_\infty^2=\frac{1}{1-\frac{2\beta \gamma}{2-\gamma}},
\]
provided $\beta<\beta_{crit}\equiv\frac{1}{\gamma}-\frac{1}{2}$. Thus, in this case, the conditions required for 
validity of the central 
limit theorem hold at sufficiently high temperature: the variables are strongly 
correlated at short scale, but become 
weakly correlated at sufficiently large scales, ensuring that the asymptotic distribution is still Gaussian, with a 
suitable variance.

Quite interestingly, at $\beta=\beta_{crit}$, the variance diverges and the asymptotic distribution for the block 
variables (\ref{block1}) can no longer be Gaussian, signaling that the variables are correlated at all scales of the 
hierarchy. This is exactly what happens in a physical system close to a critical point. To obtain a well-defined
limiting distribution, a different strategy to regroup the variables, and a different limit theorem, 
must be used. Let us
introduce the new {\em block variables}
\begin{equation}
x^{(k)}=\sum_{i=1}^{2^k}\frac{x_i}{2^k\gamma^{-k/2}},
\label{block2}
\end{equation}
different form Eq. (\ref{block1}).
Again, the asymptotic distribution for these variables can be worked out analytically. The result is that for 
$\sqrt{2}<\gamma<2$
the distribution is Gaussian with variance
\[
\sigma_\infty^2=\frac{2-\gamma}{2\gamma\beta_c}.
\]
Even more interesting is the result for $1<\gamma<\sqrt{2}$. Things become more intricate, and 
a simple solution can only be found for small $\epsilon\equiv \sqrt{2}-\gamma>0$. In this case the asymptotic 
distribution of the block variables (\ref{block2}) is non Gaussian,
\[
p_\infty(x)\approx A\,\mathrm e^{-r^*(\epsilon)x^2/2-u^*(\epsilon)x^4/4},
\]
where $A$ is a normalization constant and the parameters $r^*(\epsilon)$ and $u^*(\epsilon)>0$ 
characterize the distribution 
and are called fixed-point couplings.

Thus, the hierarchical model (\ref{hier}) is rich enough to provide a realization of nontrivial 
limit theorems embodying 
the universal behavior of a system near a phase transition.

Let us conclude this section stressing the deep link between the renormalization group procedure 
and a generalization of the central limit theorem. There is an interesting class of stable 
distribution laws of random 
variables with the following property:
let $P_\lambda(x)$ be the probability of the variable $x$ depending on the set of parameters 
$\lambda$ (e.g., in a Gaussian
distribution the $\lambda$'s amount to the mean value and the variance); denoting with $*$ 
the convolution, we have that
for the stable distribution law
\[
\left(P_{\lambda_1}*P_{\lambda_2}\right)(x)=P_\lambda(x),
\]  
where $\lambda$ is an appropriate function of $\lambda_1$ and $\lambda_2$. For instance, in the case of a Gaussian 
distribution $\lambda=(m,\sigma^2)$ with $m=m_1+m_2$ and $\sigma^2=\sigma_1^2+\sigma_2^2$. It is remarkable that the
renormalization group can be seen as a convolution with the rule
\[
P_{\lambda_n}(x)=P_{\lambda_{n-1}}(\alpha_{n-1}x)*P_{\lambda_{n-1}}(\alpha_{n-1}x),
\]
and one has a fixed point as $n\to\infty$ for a suitable choice of $\lambda_n$ and $\alpha_n$.

The simplest Gaussian fixed point (for simplicity we take $m=0$) corresponds to $\alpha_n=1/\sqrt{2}$
and $\sigma_n^2=\sigma_{n-1}^2$.

\section{Discussion}

The present Section is devoted to the discussion of a some technical, as well as conceptual, aspects  
of the probability in statistical physics
which are often underestimated.

\subsection{The maximum entropy principle}

We wish to add some remarks about information and the ``it from bit'' approach. The question here is whether 
statistical mechanics can be seen as a form of statistical inference. 
According to a radically anti-dynamical point of view,
this is indeed the case: statistical mechanics is not a theory of objective physical reality, 
and the probabilities measure the
``degree of truth'' of a logical proposition. In this context, Jaynes \cite{diciotto} proposed 
the maximum entropy principle (MEP) as a
general rule to infer the probability of a given event when only partial information is available. 
Let us briefly summarize the MEP approach:
the mean values of
$M$ independent functions $f_k(\mathbf X)$ are known,
\begin{equation}
c_k=\langle f_k\rangle=\int f_k(\mathbf X)\rho(\mathbf X)\,\mathrm d\mathbf X, ~~~~~~~ 
\mathrm{for}~~k=1,...,M\label{contraints}
\end{equation}
then the MEP rule determines the probability density $\rho(\mathbf X)$  maximizing the entropy
\[
\mathcal H=-\int \rho(\mathbf X)\log \rho(\mathbf X)\,\mathrm d\mathbf X,
\]
with   the constraints (\ref{contraints}).
Using the method of Lagrange multipliers, it  is then easy to show that
\[
\rho(\mathbf X)=\mathrm{Const.}\,\exp\left[\sum_{k=1}^M \lambda_k f_k(\mathbf X)\right],
\]
where $\lambda_1,...\lambda_M$ depend on $c_1,...,c_M$. When applied to the statistical 
mechanics of systems with fixed 
number of particles and the only constraint of a given mean value $E$ for the energy, the MEP recipe leads to the
canonical distribution in a very simple way. 
\\
As far as we know this is the unique relevant success of MEP in physics.
Up to now there is not any convincing evidence of the possibility to use
the MEP  to derive unknown results, e.g., in non equilibrium statistical mechanics \cite{auletta};
in the following we will discuss the reason of this difficulty. 

\subsection{About probability and tipicality}

Let us briefly comment upon the fact that the approach in terms of ergodicity and that 
based on MEP reflect deep conceptual differences about how to consider probability.
It is well known that there are several competing interpretations of the actual meaning 
of probability \cite{diciannove}. 
It is rather difficult to discuss in detail the different interpretations,
here our interest is limited to the probability in statistical mechanics, in particular to
the link between probability and real world.
Our main aim is to 
investigate the following topic: what is the link between the probabilistic computations 
and the actual results obtained in laboratory experiments.
Roughly speaking the different points of view  about probability
can be classifies in two large classes:
the subjective interpretation and the objective one. 
  
According to the subjective interpretation probability is a degree of belief; 
one of the most influential supporter of such a point of view is Jaynes with
his idea that the theoretical description
of physical systems is governed by the degree of belief of the observer. 

On the contrary, in the objective interpretation the probability of an event is determined by the physics of the 
systems and not by the lack of information of the observer.
In particular ergodic theory, somehow, justifies a frequentist interpretation of probability, according to which
it is possible to obtain an empirical notion of probability which is an objective property of the 
trajectory \cite{cinque,sette,venti}.
There is no universal agreement on this issue; for instance, Popper believed that probabilistic concepts are 
extraneous to a deterministic description of the world, while Einstein held the opposite 
point of view \cite{ventuno}.

Let us try to give an answer to the following bold question: 
what is the link between the
probabilistic computations (i.e., the averages over an ensemble) and
the  results obtained in laboratory experiments which, a
fortiori, are conducted on a single realization (or sample) of the
system under investigation?
In a nutshell, following Boltzmann, we can invoke
the notion of typicality, i.e., the fact that the outcome of an
experiment on a macroscopic system takes a specific (typical) value
overwhelmingly often \cite{ventidue,ventitre,ventiquattro}. 
In statistical mechanics typicality holds for $N \gg 1$. 
The concept of typicality is at the basis of the
very possibility to have reproducibility of results in experiments (on
macroscopic objects) or the possibility to have macroscopic laws.

From
the limit theorems (as well as the typicality) one has that the true deep
role of probability in statistical mechanics is the possibility to identify the mean value with
the actual result for a unique (large) system \cite{ventidue,ventitre,ventiquattro}.

The practical relevance of the typicality can be appreciated looking carefully at 
the most popular computational methods in statistical mechanics,
i.e. molecular dynamics and Monte Carlo.
The first technique is based on the assumption of the validity of
the ergodicity.
As already discussed, strictly speaking, such an assumption is surely wrong.
In the Monte Carlo computations one uses suitable ergodic Markov chains, therefore
the validity of the ergodicity is sure.
On the other hand it is easy to realize that
the success of the method cannot be based only on such a mathematical base.
From Poincar\'e's theorem, and Kac's lemma on the mean recurrence time,
one can understand \cite{sei} that for some observables in order to obtain the proper
mean value from a time average, a gigantic number of steps is necessary.
Namely a time $O(C^N)$ where $C>1$ and $N$ is the number of degrees of freedom \cite{24b}.
\\
Therefore the true reason of the success cannot be the mathematical ergodicity,
but  the typicality and it is, somehow, related to the Khinchin's results, i.e.
\\
$\bullet$ the investigated system has a huge number of degrees of freedom: $N \gg 1$;
\\
$\bullet$ in a Monte Carlo computation, usually one computes only
the average of a few observables of physical interest which involve
many degrees of freedom and therefore it is not necessary to explore
the whole phase space.

\subsection{Is then the MEP a cornucopia, or a Pandora's box?}

We raise here two main objections. The first comes from the 
old good principle {\em Ex nihilo nihil} (Nothing from nothing).

In his nice book about statistical mechanics \cite{venticinque} Ma asks: ``How many days in a 
year does it rain in Hsinchu
(a city in northern Taiwan, commonly nicknamed {\em The Windy City}, because of its windy climate)?''.
According to MEP the reply might
be: ``As there are two possibilities, to rain or not to rain, and we are completely 
ignorant about Hsinchu, it rains six 
months a year''. 
We share Ma's opinion that the above answer is simply nonsense:
it is not possible to infer something about a real phenomenon, only based on our ignorance.

But there is a more technical aspect that makes the MEP approach rather weak, namely the dependence of the results
on the choice of the variables: the MEP gives different solutions if different variables are adopted to describe the 
same phenomenon. This is due to the fact that the ``entropy'' 
\[
\mathcal H_{\mathcal X}=-\int \rho_{\mathcal X}(\mathbf X)\log \rho_{\mathcal X}(\mathbf X)\,\mathrm d\mathbf X
\]
is not an intrinsic quantity but it depends by the variable  $\mathcal X$ used for the state of the system.
Using a different parametrization,
i.e., the coordinates given by an invertible map $\mathbf Y=\mathbf f(\mathbf X)$, 
the entropy of the same phenomenon
would be
\begin{eqnarray*}
\mathcal H_{\mathcal Y}&=&-\int \rho_{\mathcal Y}(\mathbf Y)\log \rho_{\mathcal Y}(\mathbf Y)\,\mathrm d\mathbf Y
\nonumber\\
&=&\mathcal H_{\mathcal X}+\int \rho_{\mathcal X}(\mathbf X)\log 
\left|D\mathbf f(\mathbf X)\right|\,\mathrm d\mathbf X,
\nonumber
\end{eqnarray*}
where $|D\mathbf f(\mathbf X)|$ is the determinant of the Jacobian matrix, measuring 
how the elementary volume changes 
with the change of variables.

Let us now reconsider the fact that using MEP is easy to obtain the canonical distribution.
The supporters of MEP consider this result an important  success of the method.
On the other hand there is clearly a
negative aspect:  the correct result is obtained only
provided we make use of the canonical variables, i.e., positions and momenta of the particles. The choice of
the specific canonical variables is not unique, but different choices are related to one another 
by canonical transformations
that preserve the Hamiltonian structure of the equations of motion and enjoy the property that the corresponding
determinant of the Jacobian matrix equals unity.
On the contrary using different (non canonical) variables, instead of the 
$\mathrm{Const.}\, \mathrm e^{-\beta H}$ we 
have $g({\bf Y}) e^{-\beta H({\bf Y})}$ where the shape of the function $g(\,\,)$ depends on the chosen variables.

\section{Final Remarks}

Let us conclude the paper with some general comments and remarks.

Wheeler's slogan ``There is no law except the law that there is no law'' sometimes can be interpreted 
in a precise way:
In probability theory there exist limit theorems (law of large numbers, central limit theorems) which allow for
the understanding of the behavior of physical systems with many degrees of freedom.
Some words of caution are needed here. 
Sometime it is not possible to
avoid some details from fundamental theories, for instance, in the case of the Boltzmann-Gibbs 
canonical distribution.
In such a case the basic  assumption is the validity of the 
microcanonical distribution which can be  justified by a dynamical argument, namely the Liouville theorem and the
assumption of (a weak form of) ergodicity.

It is dangerous to believe too much in {\em general principles} (like MEP),
whose results, sometimes, can be used to simplify a description {\em a posteriori}, i.e., 
only after a deeper understanding 
of a given piece of physical reality by means of a more robust mathematical approach. An 
example of the above warning 
is provided by Ulam's conjecture on Boltzmann's law, which can be obtained with a stochastic 
rule in the binary collision 
process. A careful analysis shows how the correct result is obtained only with a precise 
collision rule, and therefore it
cannot be considered just the gift of a general regulatory principle.

To conclude, we stress again the basic role of the limit theorems in statistical mechanics. The law of large numbers
and the typicality are the two relevant ingredients which allow for a conceptual (and technical) link between
the use of probability and the observations in a unique (large) system.

\begin{acknowledgements}
We wish to thank M. Falcioni and A. Puglisi for many useful comments.
\end{acknowledgements}

\end{document}